# CMOS-compatible graphene photodetector covering all optical communication bands

Andreas Pospischil[1], Markus Humer[2], Marco M. Furchi[1], Dominic Bachmann[1], Romain Guider[2], Thomas Fromherz[2], and Thomas Mueller[1,*]

[1]Vienna University of Technology, Institute of Photonics,
Gußhausstraße 27-29, 1040 Vienna, Austria

[2]Johannes Kepler University Linz, Institut für Halbleiter und Festkörperphysik
Altenbergerstraße 69, 4040 Linz, Austria

**Optical interconnects are becoming attractive alternatives to electrical wiring in intra- and inter-chip communication links. Particularly, the integration with silicon complementary metal-oxide-semiconductor (CMOS) technology has received considerable interest due to the ability of cost-effective integration of electronics and optics on a single chip[1]. While silicon enables the realization of optical waveguides[2] and passive components[3], the integration of another, optically absorbing, material is required for photodetection. Germanium[4] or compound semiconductors[5] are traditionally used for this purpose; their integration with silicon technology, however, faces major challenges. Recently, graphene[6] has emerged as a viable alternative for optoelectronic applications[7], including photodetection[8]. Here, we demonstrate an ultra-wideband CMOS-compatible photodetector based on graphene. We achieve multi-gigahertz operation over all fiber-optic**

---

[*]Email: thomas.mueller@tuwien.ac.at

**telecommunication bands, beyond the wavelength range of strained germanium photodetectors[9], whose responsivity is limited by their bandgap. Our work complements the recent demonstration of a CMOS-integrated graphene electro-optical modulator[10], paving the way for carbon-based optical interconnects.**

Graphene's suitability as a photodetector was first demonstrated by locally illuminating the vicinity of one of the electrical contacts of a back-gated graphene transistor[11–13]. A detectable current was generated which was attributed to band bending at the metal/graphene interface. Photocurrents have also been obtained at single/bilayer graphene interfaces[14] and graphene p-n junctions[15] and attributed to the thermoelectric effect. The detailed microscopic processes involved in the photocurrent generation are still debated[16–18]. Further optimization of the device geometry led to high-speed electrical measurements on graphene-based photodetectors, operating at frequencies[19] >40 GHz and supporting data rates of 10 Gbit/s in an optical communication link[8]. More recent work has concentrated on increasing the responsivity of the devices[20–23] and extending their operation range to longer wavelengths[24, 25]. The research on photodetectors has been paralleled by work on graphene-based optical modulators[10, 26–28], which culminated in the demonstration of waveguide-integrated devices with >1 GHz bandwidth, operating over a broad range of wavelengths, and exhibiting device foot-prints of 25 $\mu m^2$ only[10, 26]. In these devices, electrically controlled absorption is achieved by tuning the Fermi level of a graphene sheet with a gate electrode[29, 30].

Here, we report the design and realization of a silicon waveguide-integrated graphene photodetector. The device structure is similar to that of the optical modulator recently demonstrated by[10] Liu *et al.*, though with different electrode arrangement. The integration of both devices on the same chip will in the future allow for the development of scalable ultrahigh-bandwidth graphene-based optical interconnects. In fact, the application potential of graphene for optical interconnects has recently been recognized[31].

A scanning electron microscope (SEM) image of one of our devices is shown in Figure 1 (a). An enlarged view of the section highlighted by the dashed rectangle is shown in Figure 1 (b). The devices are fabricated in a simple three-step process: (i) etching and passivation of the silicon waveguide; (ii) deposition and structuring of graphene; (iii) metallization. The details are described in the Methods section.

In a device of proper length $L$, the optical mode is almost completely absorbed as the light propagates along the silicon waveguide. The local potential gradient at the interface between the central Ti/Au electrode (signal electrode; S) and graphene drives a photocurrent PC (by photovoltaic[11–13] and/or thermoelectric[14–16] effects) towards the ground (GND) lead, as schematically illustrated in Figure 1 (c). The potential gradient originates from different dopings in the metal-covered (p-doped[11]) and uncovered ($p^+$-doped due to adsorbates from the ambient air) parts of graphene and could additionally be enhanced by utilizing the waveguide itself as a back-gate electrode to modulate the potential in the graphene channel. A ground-signal-ground (GND-S-GND) configuration is used, which allows doubling of the total photocurrent as compared to the simple GND-S case[19]. Due to the lack of an electronic bandgap in graphene, the photo-generated carriers pass through the potential barriers at the GND-electrodes almost unimpeded, leading to high-bandwidth photodetection even without S-GND bias.

However, the metallic electrode S, in close proximity to the waveguide, induces light absorption that can significantly degrade device performance. To evaluate the tradeoff in photo-responsivity, mode propagation calculations were performed using the finite element method. In our calculations, graphene is described by a complex refractive index $n(\lambda) = 3.0 + i(C_1/3)\lambda$, where $C_1$ = 5.446 µm$^{-1}$ and $\lambda$ = 1550 nm is the wavelength[32]. Only the in-plane (tangential) electric field component is considered when calculating the optical absorption. The other materials (Si, SiO$_2$, Ti, and Au) are modeled by their respective wavelength-dependent complex refractive indices. Since scattering can be neglected[33] (because of the strong mode confinement and the uniform device structure), the mode damping is only dependent on $\alpha_M$ and $\alpha_G$, which represent the absorption coefficients of the metal electrode and graphene, respectively. The device geometry in the simulations corresponds to the one shown in the inset of Figure 1 (a).

First, in order to obtain $\alpha_M$, calculations were carried out in which the graphene absorption was set to zero ($C_1$ = 0). Calculations were performed for different widths $W$ of the signal electrode and are shown (symbols) for the fundamental quasi-TE mode in Figure 2 (a). As expected, $\alpha_M$ strongly increases with increasing electrode width. If half the waveguide is covered with metal, the propagation length of light is reduced to 7 µm only. In addition to the fundamental TE mode, the waveguide also supports the propagation of the lowest-order quasi-TM mode, whose damping $\alpha_M$, however, is significantly larger. We thus expect the photodetector to be polarization-sensitive, with higher responsivity for TE polarized light. In a second step, we calculate $\alpha_G$ by setting $C_1$ ≠ 0 and integrating the tangential field-component along the graphene sheet. The results for both monolayer and bilayer graphene are shown as dashed and solid lines, respectively, in Figure 2 (a). For $W$ > 100 nm (monolayer) and $W$ > 160 nm (bilayer), the absorption of light by the metal electrode is more dominant than the absorption in graphene. For achieving high responsivity, it is thus necessary to keep the electrode width *W* as small as possible. On the other hand, the contact resistance increases with decreasing *W*, which decreases the photocurrent and gives rise to reduced RC-bandwidth. These requirements contradict each other, and the device parameters have to be optimized by minimizing the trade-offs.

With $\alpha_M$ and $\alpha_G$, the fraction $\eta$ of light that is absorbed in the graphene sheet (see Figure 2 (b)) can be calculated by $\eta(L) = \frac{\alpha_G}{\alpha_G + \alpha_M}\left(1 - e^{-\alpha_G L}e^{-\alpha_M L}\right)$. The calculation results show that efficient light absorption can be achieved with short device lengths, enabling high-speed operation and dense integration capability. For example, $\eta > 50\ \%$ is obtained in an only 22 μm long bilayer device (*W* = 100 nm). Moreover, as can be seen from the calculated mode profile ($E_x$-component) in Figure 2 (a), the absorption occurs mainly near the metal/graphene interface where photocurrent generation is most efficient. Note that light ($\approx 34\ \%$ in the above example) is also absorbed in the graphene covering the sidewalls of the waveguide [mainly caused by the $E_z$-component (not shown)].

We will now present performance characteristics of a bilayer graphene device with *L* = 24 μm and *W* = 180 nm. For this device dimensions, we estimate $\eta \approx 44\ \%$ according to Figure 2 (b). For device characterization, laser light with 1550 nm wavelength was coupled into the silicon waveguide using a lensed single-mode fiber ($\approx 2.5$ μm spot diameter). The optical power at the input port of the waveguide-integrated graphene photodetector, $P_{in}$, was estimated from transmission measurements of reference waveguides without photodetector. The total losses, 15±1 dB, mainly stem from coupling losses from the optical fiber to the waveguide. A similar value is obtained by calculating the overlap integral between the field profile of the incident wave and that of the guided mode. The device was connected to a transimpedance amplifier with low input impedance to measure the photocurrent $I_{pc}$. Figure 3 (a) shows the power dependence of $I_{pc}$, which is linear across the entire measurement range. The photo-responsivity $S$, defined as the ratio of the measured photocurrent to the input optical power, is $S = \frac{I_{pc}}{P_{in}} \approx 0.03$ AW$^{-1}$. Our best device, made of tri-layer graphene, had a responsivity of $S \approx 0.05$ AW$^{-1}$ – an order of magnitude larger than that achieved with normal-incidence graphene photodetectors[8]. From the 0.05 AW$^{-1}$ we estimate an internal quantum efficiency of $\xi = S\frac{hc}{e\lambda\eta} \approx 10\%$ (Planck's constant, $h$; speed of light, $c$; electron charge, $e$). Considering the $\xi$ = 30–60 % that have been estimated from photocurrent measurements on simple metal/graphene interfaces[13, 34], we believe that there is plenty of room for improvement – for example by applying gate[11–13] or S-GND bias fields[8], improving the material quality[16], increasing the steepness of the potential gradient by using Ag instead of Au electrodes[35], increasing $\eta$ through an optimized electrode arrangement, employing graphene p-n junctions[15, 16], or suspending the graphene sheet in air[34].

The wavelength-dependence of the photo-response, shown in Figure 3 (b), was measured using three separate light sources: fixed wavelength lasers, operating at 1310 nm (O-band) and 1650 nm (U-band), respectively; and a tunable laser, operating in the range 1550–1630 nm (from the C-band, all over the L-band, to the U-band). The responsivity is flat across all optical telecommunication windows, unlike the drastic decrease of the response of Ge photodetectors beyond[1] 1550 nm, or strained Ge detectors beyond[9] 1605 nm. We expect the device to work at even longer wavelengths; limited only by the cut-off properties of the silicon waveguide. Because InGaAs cannot be monolithically integrated with silicon CMOS, other materials are currently being investigated for photodetection in the L- and U-bands, with ion-implanted Si[36] and GeSn[37] being the most promising ones. However, implanted Si detectors suffer from low optical absorption requiring millimeter long devices to absorb the radiation, resulting in device footprints that are 10–100 times larger than what is presented here. GeSn photodetectors, on the other hand, exhibit high dark currents and waveguide-integration of GeSn has, to our knowledge, not yet been demonstrated.

Figure 4 shows the RF characteristics of our photodetector, as determined by an impulse response measurement. A train of 100 fs long (full width at half maximum, FWHM) optical pulses, produced by a mode-locked erbium fiber laser (1550 nm center wavelength), was coupled into the device. Dispersion in the lensed fiber broadens the optical pulses to $\approx 1$ ps, which is sufficiently short for our purposes. The impulse response of the graphene photodetector was monitored with a 20 GHz bandwidth sampling oscilloscope and is shown in the main panel. The 3 dB-bandwidth of the detector output $\Delta f$, as obtained from the commonly used optical time-bandwidth expression $\Delta f \Delta t = 0.441$ ($\Delta t$ is the FWHM pulse duration), is $\approx 18$ GHz. This is approximately also the bandwidth of our measurement system and we anticipate that the device might work at even higher frequencies. Fourier transform of the time-domain data provides the spectrum, shown in the inset.

Let us finally summarize the opportunities that graphene offers as a new material for optical interconnects: (i) *Ultra-wideband operation.* The gapless character of graphene enables optical interband transitions to occur over an ultra-wide wavelength range, unmatched by any other material. In this letter, we demonstrate photodetector operation from the O- to the U-band. However, graphene-based optoelectronic devices can operate over an even wider range of wavelengths[8, 22, 24, 25]. (ii) *High-speed operation.* With its extremely high carrier mobility – 200,000 cm$^2$ V$^{-1}$ s$^{-1}$ at room temperature – graphene is predestined for high-speed applications, including high-speed data transmission. (iii) *Low energy consumption.* Besides high data rate, low energy consumption (measured in Joules/Bit) is the most important requirement for an optical communication link. Our devices rely on built-in potentials that exist at metal/graphene interfaces and are hence operated under zero-bias conditions (i.e., without dark current) and vanishing power consumption. Moreover, it has recently been predicted that the energy consumption of graphene-based modulators can be at least as low as the best modulators fabricated using Si and SiGe technology[38] (<0.5 fJ/Bit). Graphene could hence provide substantial advantages for use in energy-efficient optical interconnects. (iv) *Small device footprint.* The strong optical absorption in graphene (absorption coefficient at infrared wavelengths more than 100 times higher than that of Ge) allows for a high level of integration of optoelectronic devices on a single chip. With device footprints down to 50 μm$^2$ (without contact pads), about 20,000 of our photodetectors could be integrated on a 1×1 cm$^2$ die. A similar integration density is predicted for graphene-based modulators[10]. (v) *Compatibility with CMOS and other technologies.* Graphene's two-dimensional character makes it compatible with standard semiconductor technology and allows for the monolithic integration with silicon and other materials. Our devices operate under voltage and current requirements of CMOS and their fabrication does not involve any high-temperature processes or contamination issues that could harm silicon circuitry. The recent developments in high-quality CVD graphene synthesis[39] will allow for wafer-scale integration of such devices into optical interconnects. The mechanical flexibility of graphene also enables the integration with bendable substrates. We envision that graphene could particularly play a role in realizing various active components in polymer-based optical circuits[40]. (vi) *Simplicity and low-cost.* The versatility and broadband capability of graphene allows for intriguingly simple device layouts and offers low development and fabrication costs by eliminating the need for multiple detector and modulator designs. All these aspects lead us to consider graphene as a promising new material for integrated photonics.

**Methods**

Our devices are fabricated on a commercial silicon-on-insulator (SOI) wafer with 270 nm thick Si device layer and 3 μm buried-oxide. A 600 nm wide waveguide was defined using optical lithography and etched by reactive-ion etching (RIE). The wafer was then covered by plasma-enhanced chemical vapor deposition (PECVD) with a 7 nm thick layer of $SiO_2$ to prevent electrical contact between the graphene and the silicon waveguide. A graphene sheet of suitable size and thickness (verified by optical microscopy and Raman spectroscopy[41]) was then prepared by mechanical exfoliation on a separate 300 nm thick oxide-coated Si wafer, lifted off from this wafer, and transferred with micrometer-precision onto the desired location on the waveguide sample. Care was taken to avoid placement of graphite chunks (which are, inevitably, also transferred in this process) on top of the optical waveguide. The graphene sheet was etched in shape by oxygen plasma (only in samples where necessary) and contact electrodes were fabricated by electron-beam lithography and metal deposition. The metal deposition occurred in two steps: First, a nominally 1 nm thick titanium (Ti) layer was evaporated, followed by a 20 nm thick gold (Au) layer. The Ti serves to improve the adhesion of Au to graphene and does not form a continuous layer. In a second step, a 25 nm thick Au layer was sputtered to ensure that the waveguide sidewalls are covered with metal in order to achieve electrical contact between the central electrode and the bonding pad. Keeping the Ti content as low as possible is crucial, as its dissipative dielectric function leads to a strong damping of the optical mode. The sample was annealed in vacuum at 125 °C for several hours to remove PMMA residues from the surface of the graphene. Finally, the sample was cleaved using a diamond scribe in order to obtain a clean facet for the in-coupling of light. A schematic of the device cross-section is shown in the inset of Figure 1 (a).

**Acknowledgements**

We acknowledge helpful discussions with Werner Schrenk and Karl Unterrainer. We are very grateful to Holger Arthaber for lending us a sampling oscilloscope and to Karl Unterrainer, Juraj Darmo, and Daniel Dietze for providing us access to their femtosecond fiber laser. This work was supported by the Austrian Science Fund FWF (START Y-539) and the Austrian Research Promotion Agency FFG (NIL-Graphene, PLATON-SiN).

**Author contributions**

T.M. conceived and designed the experiments. A.P. and M.H. fabricated the samples. A.P. carried out the measurements. T.M. and M.H. performed the simulations. M.F. and

**Additional information**

**Competing financial interests**

The authors declare that they have no competing financial interests.

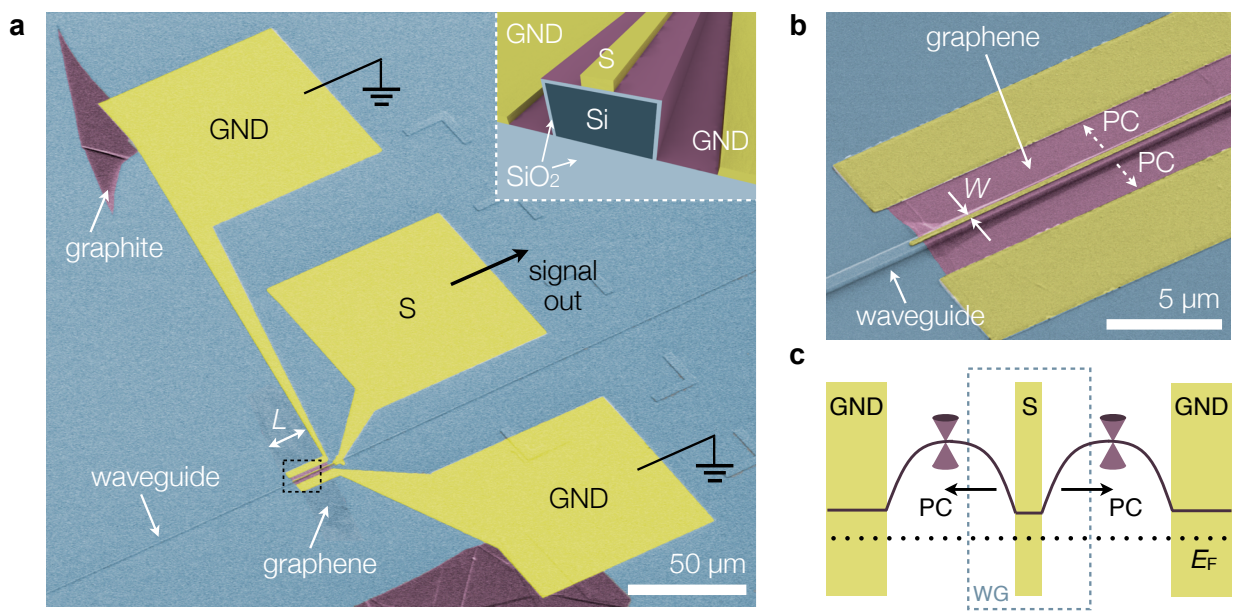

*Figure 1.* **Waveguide-integrated graphene photodetector.** (a) Colored scanning electron micrograph of a waveguide-integrated graphene photodetector. The "active region" of the graphene sheet is shown in violet. Scale bar, 50 μm. Inset: Cross-section of the device. The graphene sheet coats both the top surface and sidewalls of the waveguide. A thin $SiO_2$ layer prevents electrical contact between graphene and the silicon waveguide. (b) Enlarged view of the section highlighted by the dashed rectangle in (a). The photocurrent (PC) flows from the central signal electrode (S) towards the ground (GND) electrodes. Scale bar, 5 μm. (c) Schematic illustration of the band diagram. The dotted line represents the Fermi level $E_F$. Photocurrent (PC) is generated in the vicinity of the signal electrode (S).

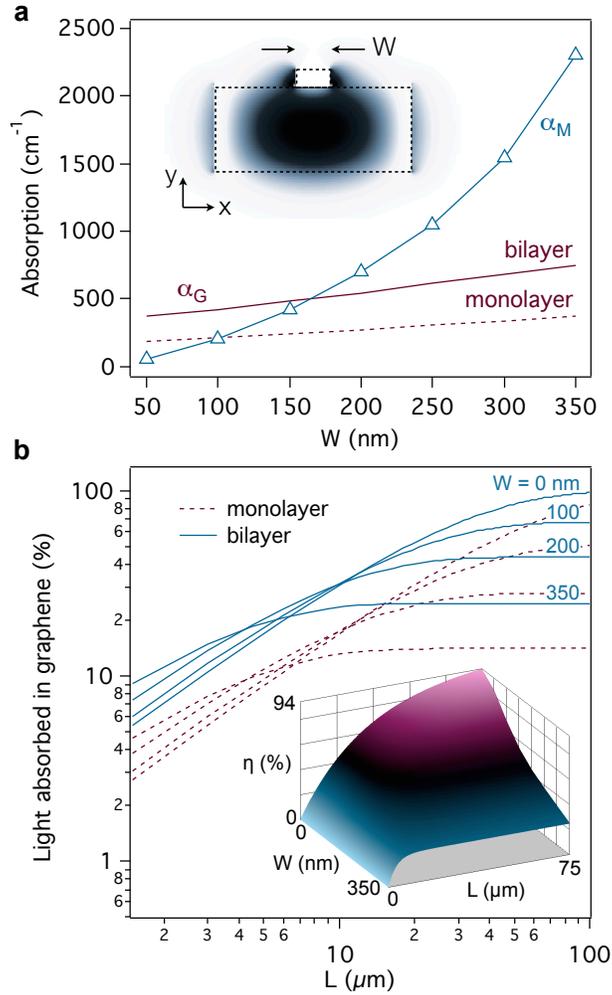

*Figure 2.* **Simulation results and device optimization.** (a) Comparison of the losses in the metallic signal electrode $\alpha_M$ (symbols) and the graphene absorption $\alpha_G$ (dashed line: monolayer graphene, solid line: bilayer graphene) as a function of electrode width $W$. The calculations were performed for 1550 nm wavelength and TE polarization. For $W > 100$ nm (monolayer) and $W > 160$ nm (bilayer), the losses are more dominant than the absorption in graphene. Inset: Mode profile ($E_x$-component) of the fundamental quasi-TE mode. The contours of the waveguide and the signal electrode are shown as dashed lines. (b) Light absorption in the graphene sheet as a function of device length $L$ and electrode width $W$ (plotted on a double-logarithmic scale). Curves are presented for $W = 0$ nm (no electrode), $W = 100$ nm, $W = 200$ nm, and $W = 350$ nm. Solid lines: bilayer graphene; dashed lines: monolayer graphene. The inset shows the same data for bilayer graphene on a linear scale. For $W = 0$, the absorption in graphene can reach 100 % if the device is made sufficiently long. For $W \neq 0$, losses in the metal electrode restrict the maximum achievable absorption (and hence efficiency) of the device.

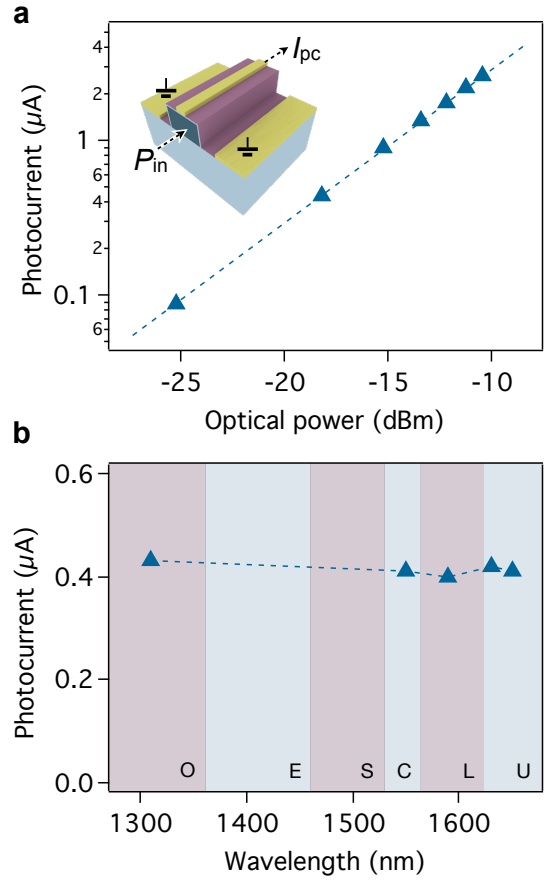

*Figure 3.* **Performance characteristics of a bilayer graphene photodetector.** (a) Photocurrent as a function of the incident optical power. The responsivity $S = I_{pc}/P_{in}$ is extracted from the slope of the curve. (b) Wavelength-dependence of the photocurrent. A flat response is obtained across all optical telecommunication windows.

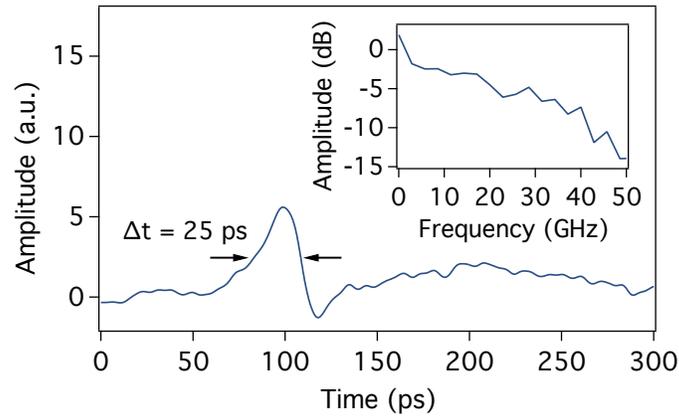

*Figure 4.* **High-speed photo-response of a bilayer graphene photodetector.** Main panel: Impulse response of the device recorded with a 20 GHz sampling oscilloscope. The FWHM pulse duration is $\Delta t \approx 25$ ps, which translates into a bandwidth of $\Delta f = 0.441/\Delta t \approx 18$ GHz. Inset: The frequency response is obtained by Fourier transform of the time-domain data.